\documentclass[aps,showpacs,pre,superscriptaddress,twocolumn]{revtex4-1}
\usepackage{footmisc}
\usepackage{graphicx}
\usepackage{amssymb}
\usepackage{amsmath}
\usepackage{dsfont}
\usepackage{setspace}
\usepackage{dcolumn}
\usepackage{newfloat}
\usepackage{epsfig}
\usepackage{multibib}
\usepackage[dvipsnames]{xcolor}
\usepackage{mathtools}
\usepackage{hyperref}
\hypersetup{colorlinks=true, citecolor=Blue, linkcolor=Blue, urlcolor=Blue}

\usepackage{epstopdf}
\usepackage{bm}
\usepackage{float}
\usepackage{color}

\def\bea{\begin{eqnarray}}
\def\eea{\end{eqnarray}}
\def\be{\begin{equation}}
\def\ee{\end{equation}}

\usepackage{datetime}
\advance\currenthour by 0

\begin{document}

\title{Raman Spectroscopy and Aging of the Low-Loss Ferrimagnet Vanadium Tetracyanoethylene}

\author{H. F. H. Cheung}
\affiliation{School of Applied and Engineering Physics, Cornell University, Ithaca, NY 14853, USA}

\author{M. Chilcote}
\affiliation{School of Applied and Engineering Physics, Cornell University, Ithaca, NY 14853, USA}

\author{H. Yusuf}
\affiliation{Department of Physics, The Ohio State University, Columbus, Ohio 43210, USA}

\author{D. S. Cormode}
\affiliation{Department of Physics, The Ohio State University, Columbus, Ohio 43210, USA}

\author{Y. Shi}
\affiliation{Department of Physics and Astronomy, University of Iowa, Iowa City, Iowa 52242, USA}

\author{M. E. Flatté}
\affiliation{Department of Physics and Astronomy, University of Iowa, Iowa City, Iowa 52242, USA}

\author{E. Johnston-Halperin}
\affiliation{Department of Physics, The Ohio State University, Columbus, Ohio 43210, USA}

\author{G. D. Fuchs}
\email[Corresponding author: ]{gdf9@cornell.edu}

\affiliation{School of Applied and Engineering Physics, Cornell University, Ithaca, NY 14853, USA}
\affiliation{Kavli Institute at Cornell for Nanoscale Science, Ithaca, New York 14853, USA}

\begin{abstract}
Vanadium tetracyanoethylene (V[TCNE]$_{x}$, $x\approx 2$) is an organic-based ferrimagnet with a high magnetic ordering temperature $\mathrm{T_C>600 ~K}$, low magnetic damping, and growth compatibility with a wide variety of substrates. However, similar to other organic-based materials, it is sensitive to air. Although encapsulation of V[TCNE]$_{x}$ with glass and epoxy extends the film lifetime from an hour to a few weeks, what is limiting its lifetime remains poorly understood. Here we characterize encapsulated V[TCNE]$_{x}$ films using confocal microscopy, Raman spectroscopy, ferromagnetic resonance and SQUID magnetometry. We identify the relevant features in the Raman spectra in agreement with \textit{ab initio} theory, reproducing $\mathrm{C=C,C\equiv N}$ vibrational modes. We correlate changes in the effective dynamic magnetization with changes in Raman intensity and in photoluminescence. Based on changes in Raman spectra, we hypothesize possible structural changes and aging mechanisms in V[TCNE]$_x$. These findings enable a local optical probe of V[TCNE]$_{x}$ film quality, which is invaluable in experiments where assessing film quality with local magnetic characterization is not possible.
\end{abstract}

\maketitle
\section{Introduction}

In the field of coherent magnonics, where function is derived through the creation and manipulation of long-lived spin wave modes, yttrium iron garnet (YIG) is the prototype material due to its exceptionally low damping at room temperature. However, it is difficult to integrate with other materials, requiring deposition on a lattice matched substrate, e.g. gadolinium gallium garnet, and a synthesis temperature above 800$^{\circ}$C \cite{Hauser2016,Sun2012,Liu2014}. In comparison, vanadium tetracyanoethylene (V[TCNE]$_{x}$, $x\approx 2$) is a low loss organic-based material with intrinsic damping comparable to YIG, and it can be grown as a film with chemical vapor deposition at a mild temperature (50$^{\circ}$C) and on nearly arbitrary substrates \cite{Yu2014,Zhu2016}. Furthermore, V[TCNE]$_{x}$ can be patterned using electron-beam lithography lift-off techniques while retaining its low magnetic loss \cite{Franson2019}. Hence it is an attractive  alternative for magnonics. In particular, it is a candidate for coupling spin wave modes to isolated defect centers~\cite{vanderSar2015,Andrich2017,Candido2020,Bertelli2020}. The ability to shape magnon modes with small volumes allows the realization of strong defect spin-magnon coupling~\cite{Candido2020}.

A practical challenge in working with V[TCNE]$_{x}$ is that, like many organic-based and monolayer materials, it is air sensitive. The recent development of encapsulation techniques, similar to those used to protect organic light emitting diodes (OLEDs) \cite{Choi2008,Chwang2003,Subbarao2010,Grover2011}, can extend its lifetime at room temperature and in ambient atmosphere from an hour to a few weeks \cite{Froning2015}. However, the limitations in the lifetime of V[TCNE]$_{x}$ and its associated aging mechanisms are not well understood. For example, one aging mechanism is the chemical reaction with oxygen and water, which turns a V[TCNE]$_{x}$ film from blue-green opaque to transparent \cite{Froning2015}. This is not the only mechanism, as evidenced by aging observed in samples stored in an argon glove box and the slow change in the magnetic properties of encapsulated samples that remain opaque. Another piece of evidence is the slowing down of aging when the sample is stored at low temperature (-30$^{\mathrm{\circ}}$C) in argon, suggesting an internal change being the next dominant aging mechanism.

First we study accelerated aging through laser heating, which provides a clean condition where reactions with oxygen and water are absent. Studying sample response to high intensity laser illumination is also of particular relevance to proposed quantum interfaces between spin waves and isolated defects, which are probed with a focused laser beam. We observe a nonlinear dependence of laser damage on optical power, which is consistent with a heating-based laser damage mechanism. Instead of being merely a nuisance, this local laser damage opens up a new avenue of V[TCNE]$_x$ patterning, which is an alternative to electron-beam patterning \cite{Franson2019}. We show preliminary results on laser patterning of V[TCNE]$_x$ and discuss its resolution limit.

To better deploy V[TCNE]$_x$ outside of a pristine environment, e.g. inert atmosphere, we need a better understanding of its aging mechanisms. V[TCNE]$_x$ is amorphous with a local structural order \cite{Haskel2004}, making Raman spectroscopy an effective method for studying its structural properties locally, at the micron scale. In addition to this optical probe, we concurrently measure magnetic properties using ferromagnetic resonance (FMR) and superconducting quantum interference device (SQUID) magnetometry to correlate changes in chemical properties with changes in magnetic properties.

The paper is structured as follows. We first describe sample fabrication and then explain the measurement procedures for confocal microscopy, micro-focused Raman spectroscopy, FMR, and SQUID magnetometry. Next, we characterize V[TCNE]$_x$ using the above methods. In particular, we show Raman spectra and explain the associated vibrational modes and their relation to chemical structure. Next we study laser damage in detail, monitoring how photoluminescence and Raman spectra change with laser exposure. We show proof-of-concept patterning examples using laser damage to selectively remove magnetism. After measuring the optical and magnetic properties of pristine V[TCNE]$_x$, we next study its aging by monitoring how the above properties change in time. In particular, we observe aged samples are more susceptible to laser damage. We also identify Raman peaks observed in pristine samples which are absent in aged or laser damaged samples. 

\section{Methods}
\subsection{Sample growth}

V[TCNE]$_{x}$ films are grown by chemical vapor deposition, similar to previous reports \cite{Yu2014}. We examine 4 samples in the main text. The first sample (sample 1) is a uniform 400 nm V[TCNE]$_{x}$ film with 73 nm of aluminum deposited with an evaporator, encapsulated with epoxy and a glass cover slip. V[TCNE]$_x$ thickness is estimated from growth time. The aluminum layer is designed to prevent the laser light from exciting epoxy, ensuring we are only probing V[TCNE]$_x$ Raman spectra and photoluminescence.

Sample 2 is a 1.6 $\mu$m thick sample. Sample 3 and 4 are 400 nm thick samples grown in the same batch. All three samples are encapsulated with glass and epoxy only.

\subsection{\textit{Ab initio} calculations}

The electronic structure and phonon modes of V[TCNE]$_x$ are calculated using the Vienna ab initio Simulation Package (VASP) (version 5.4.4) with a plane wave basis and projector-augmented-wave pseudopotentials~\cite{Kresse1993,Kresse1994,Kresse1996,KressePRB1996}. These pseudopotentials use the generalized gradient approximation (GGA) of Perdew, Burke, and Ernzerhof (PBE) \cite{Perdew1996}. A Hubbard constant U=4.19, determined via a linear response method \cite{Cococcioni2005}, was used in the phonon calculations. The hybrid functional Heyd–Scuseria–Ernzerhof (HSE) \cite{Heyd2003} with the standard range separation parameter $\omega$=0.2 was also tested for this system and showed consistent results with the PBE+U approach.

\subsection{Optical and Magnetic Measurements}
Optical measurements including micro-focused Raman spectroscopy and photoluminescence are performed in a homebuilt confocal microscope with a 532 nm continuous wave laser. Integrated light collected from the sample is detected using a single photon counting module, whereas Raman spectra are recorded using a Princeton Instruments spectrometer and low dark count camera.

We measure angle-resolved, field modulated FMR in a homebuilt spectrometer. The setup consists of a microwave signal generator, an electromagnet, a pair of modulation coils, a microwave diode detector and a lock in amplifier. Film magnetization is measured using a Quantum Design MPMS 3 SQUID magnetometer. Applied magnetic field is in plane.

\section{Raman spectroscopy}

\begin{figure}
    \centering
    \includegraphics[width=0.45\textwidth]{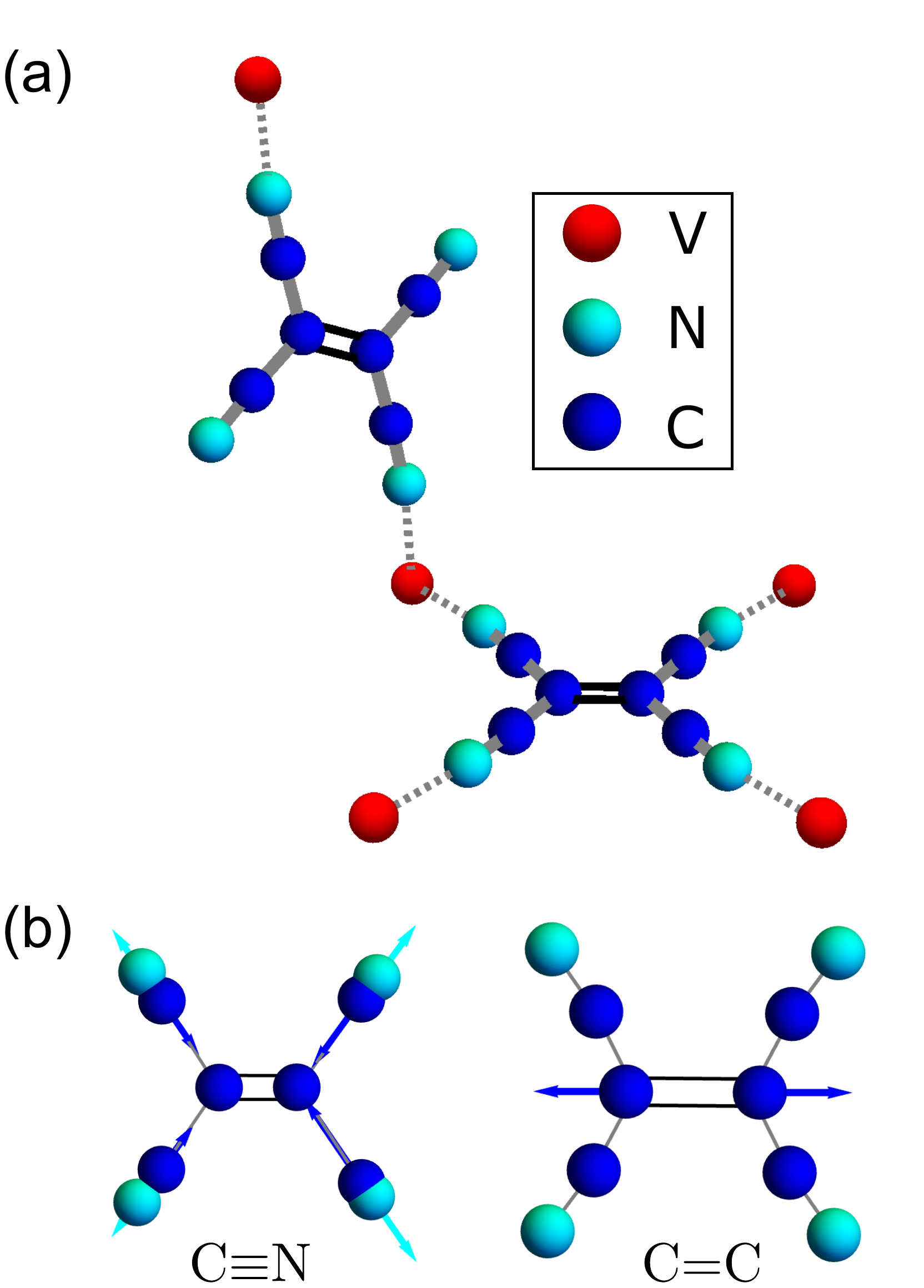}
    \caption{V[TCNE]$_x$ structure and vibrational modes based on density functional theory (DFT) calculations. (a) V[TCNE]$_{x}$ structure, showing 2 nonequivalent TCNE molecules. The bottom TCNE is bonded with 4 vanadium atoms ($\mu 4$ bonded) while the top TCNE is bonded with 2 vanadium atoms (trans-$\mu 2$ bonded) (b) $\mathrm{C\equiv N}$ (2352 $\mathrm{cm^{-1}}$) and $\mathrm{C=C}$ (1451 $\mathrm{cm^{-1}}$) vibrational modes. Shown is the bottom ($\mu$4 bonded) TCNE. High wavenumber vibrational modes ($\mathrm{>1000 cm^{-1}}$) are localized on one or the other TCNE, with non-degenerate vibrational frequencies.}
    \label{fig:VTCNE_structure}
\end{figure}

\begin{figure}
    \centering
    \includegraphics[width = 0.45\textwidth]{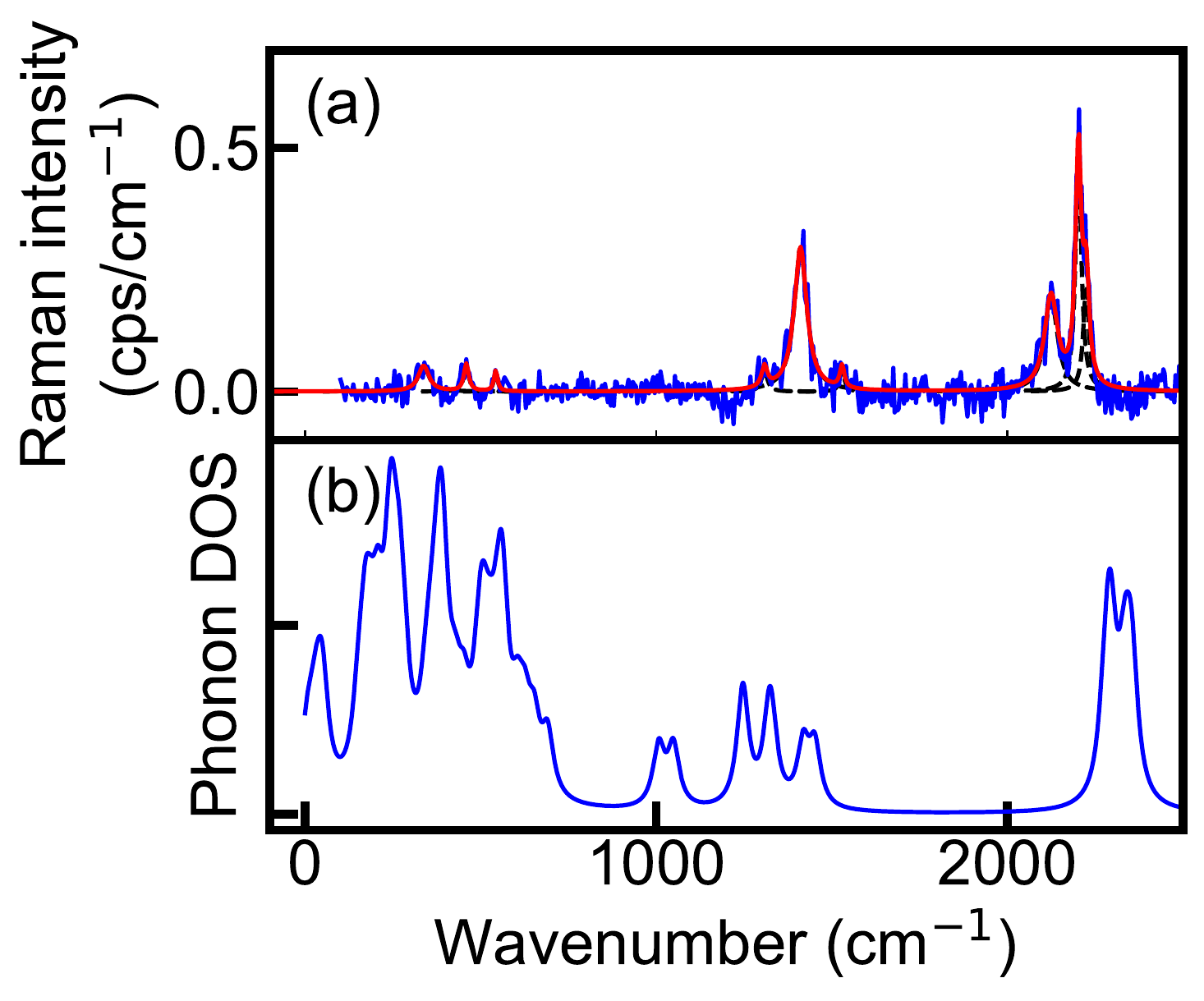}
    \caption{V[TCNE]$_{x}$ Raman spectrum and density of states (DOS). (a) Experimental spectrum of pristine encapsulated V[TCNE]$_x$. A broad baseline is subtracted from the raw Raman spectrum (see SI). (b) \textit{Ab initio} V[TCNE]$_x$ phonon DOS. In the plot, each mode is broadened as a Lorentzian with a FWHM of 20 $\mathrm{cm^{-1}}$.}
    \label{fig:VTCNE_Exp_DFT}
\end{figure}

We measure Raman spectra to characterize V[TCNE]$_x$ chemical bonds and structure. To better understand the Raman spectrum, we also perform density functional theory (DFT) calculations using VASP code. We show the calculated V[TCNE]$_x$ structure in figure \ref{fig:VTCNE_structure}(a), where the geometry agrees with previous DFT studies \cite{Fusco2009,Cimpoesu2014,Frecus2014}. In figure \ref{fig:VTCNE_Exp_DFT}, we plot the experimental Raman spectrum with DFT phonon density of states. Based on comparison with DFT calculations, we assign 1300-1500 $\mathrm{cm^{-1}}$ Raman peaks to $\mathrm{C=C}$ stretching modes, and 2200 $\mathrm{cm^{-1}}$ Raman peaks to $\mathrm{C\equiv N}$ stretching modes. Example vibrational modes are shown in figure \ref{fig:VTCNE_structure}(b), displaying modes with dominant $\mathrm{C\equiv N ~(2352~cm^{-1})}$ or $\mathrm{C=C~(1451~cm^{-1})}$ bond stretching.

Next, we explain the fine Raman features and compare them with previous studies. Fine features near 2202, 2225 $\mathrm{cm^{-1}}$ are in agreement with previous IR studies 2194, 2214 $\mathrm{cm^{-1}}$ \cite{Froning2015,Pokhodnya2000}. We observe 2121 $\mathrm{cm^{-1}}$ in Raman spectroscopy, while previous IR spectroscopy only observed a peak at 2155 $\mathrm{cm^{-1}}$ \cite{Froning2015}, which could be explained by different IR and Raman activities of these vibrational modes. The Raman peaks at 1308, 1411, 1530 $\mathrm{cm^{-1}}$ are absent in IR spectrum, suggesting vanadium atoms are more symmetrically bonded to the center $\mathrm{C=C}$ bond, leading to low IR activity \cite{Pokhodnya2000}. We attribute low wavenumber peaks at 336, 457, 543 $\mathrm{cm^{-1}}$ to low energy TCNE vibration modes and $\mathrm{V-N}$ stretching modes.

Here we have characterized Raman spectra of pristine encapsulated V[TCNE]$_x$ and assigned Raman peaks to particular vibrational modes. Low wavenumber modes are particularly relevant for magnetism as they are influenced by V--N bonds. In subsequent sections, we measure how these Raman peaks evolve under aging and laser heating induced damage, and show how Raman spectra could be used to assess film quality.

\section{Photoluminescence and laser damage susceptibility}

\begin{figure*}
    \centering
    \includegraphics[width=0.9\textwidth]{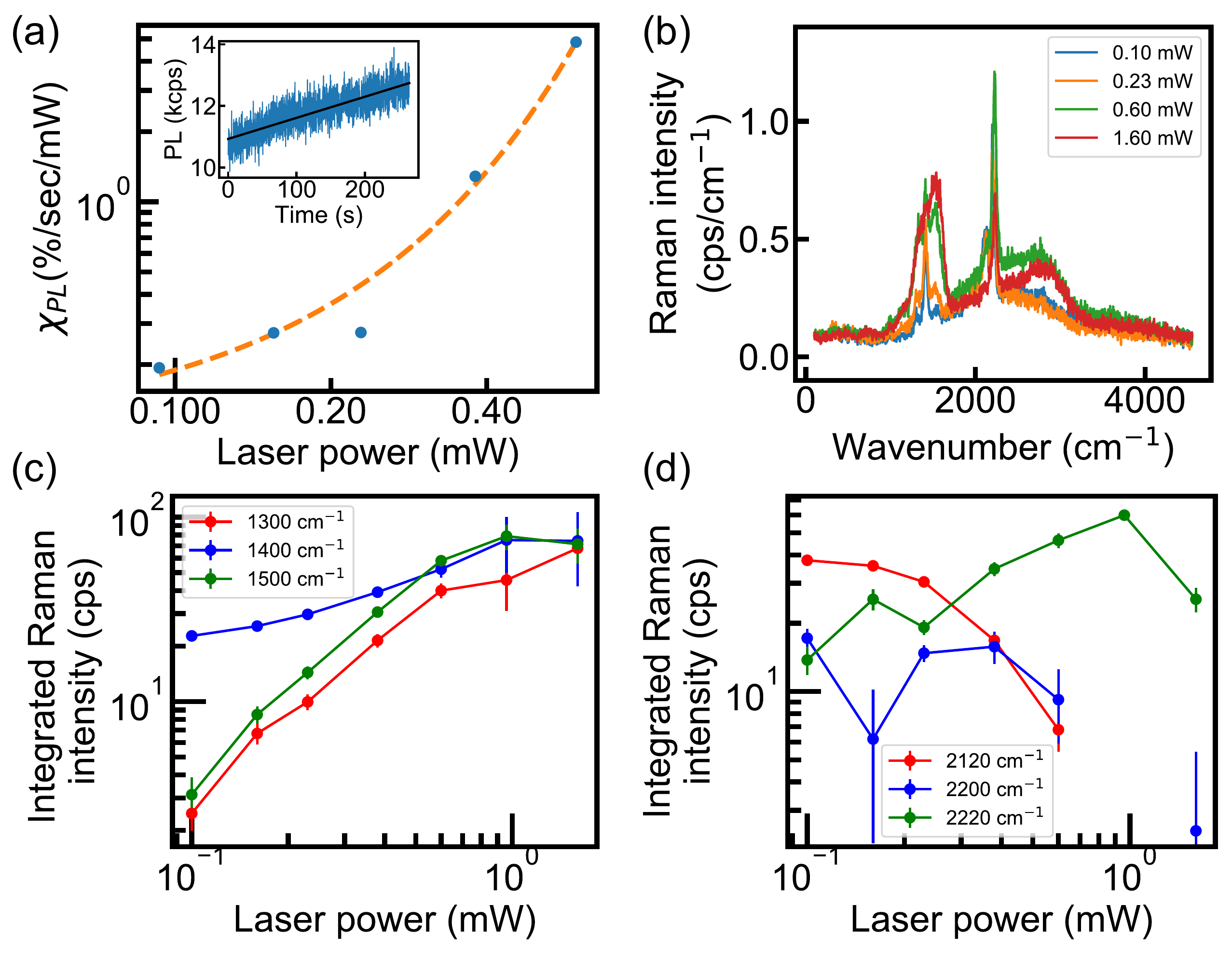}
    \caption{(a) Laser damage susceptibility $\chi_{PL}$ showing a nonlinear dependence on optical power. Also shown is a guide to the eye showing an exponential dependence with power. 
    Inset - photoluminescence time trace at an optical power 0.23 mW. (b) Raw Raman spectra taken at 0.10 mW after exposing the laser of different powers for 4 minutes. (c)-(d) Baseline subtracted integrated Raman intensity around 1400 $\mathrm{cm^{-1}}$. (d) Integrated Raman intensity around 2200 $\mathrm{cm^{-1}}$.}
    \label{fig:sample_1_laser_damage}
\end{figure*}

In the following section, we study laser damage in a pristine encapsulated V[TCNE]$_x$ film (sample 2). Using laser-induced damage allows us to study aging in the absence of oxygen and water. There is also independent interest in studying V[TCNE]$_x$ under focused laser intensity, in particular in relation to experiments coupling V[TCNE]$_x$ with defect centers \cite{Candido2020}. The goal here is to gain insights on the nature of aging and a quantitative measure of laser damage susceptibility. We note that the quantitative laser damage susceptibility is sample and encapsulation dependent and we focus on reproducible trends across samples and relative changes.

Because the laser light is tightly focused, local temperature can be large enough to cause degradation. Indeed, photoluminescence increases under continuous illumination, which is accompanied by the V[TCNE]$_x$ film turning transparent. Here we characterize laser damage susceptibility $\chi_{PL}$ as the fractional rate of change in photoluminescence per unit optical power, namely
\begin{align}
    \chi_{PL}=\frac{1}{P_{opt}}\frac{1}{PL(t=0)}\frac{d}{dt}PL(t)
\end{align}
where $PL$ is the photoluminescence, $P_{opt}$ is the incident laser power. We linear fit the initial photoluminescence increase (Fig. \ref{fig:sample_1_laser_damage}(a) inset) and extract $\chi_{PL}$.

Laser damage susceptibility increases nonlinearly in optical power, with a threshold near 0.25 mW (Fig. \ref{fig:sample_1_laser_damage}(a)). A nonlinear laser damage susceptibility is consistent with a laser heating damage mechanism. While the temperature rise is proportional to laser power, chemical reaction rates increase nonlinearly with temperature, causing more rapid aging at a high optical power.

Next we quantify how laser damage alters the Raman spectra. All measurements are done in a local area 20 $\mu$m $\times$ 20 $\mu$m, where the film is uniform. We successively illuminate different spots with different optical power for 4 minutes, and then measure Raman spectra with a low laser power (93 $\mu$W) for 4 minutes, with a peak laser intensity $\mathrm{5\times 10^4 ~W/cm^2}$. As the spot is exposed to higher laser power, the Raman intensity near 2121 $\mathrm{cm^{-1}}$ decreases with a corresponding increase in the 2202, 2225 $\mathrm{cm^{-1}}$ features (Fig. \ref{fig:sample_1_laser_damage}(b)). The side peaks at 1308, 1530 $\mathrm{cm^{-1}}$ of C=C bond increase in intensity and linewidth. The lower wavenumber peaks $\mathrm{<600~cm^{-1}}$ disappear. In addition, the total fluorescence background increases with increasing laser power. The quantitative changes in Raman intensity are shown in figure \ref{fig:sample_1_laser_damage}(c,d). Focusing on qualitative features, the disappearance of low wavenumber ($\mathrm{<600~cm^{-1}}$) and the 2121 $\mathrm{cm^{-1}}$ Raman peaks are clear, qualitative signatures of film aging. 
In particular, low wavenumber vibrational modes have significant V--N stretching components, therefore the disappearance of those peaks suggest a reduction in vanadium bonding to TCNE groups.
Likewise, the 2121 $\mathrm{cm^{-1}}$ Raman peak corresponds to a $\mathrm{C\equiv N}$ vibrational modes, which is sensitive to vanadium nitrogen bonding and is relevant for magnetic ordering in V[TCNE]$_x$. 

\begin{figure}[H]
    \centering
    \includegraphics[width=0.45\textwidth]{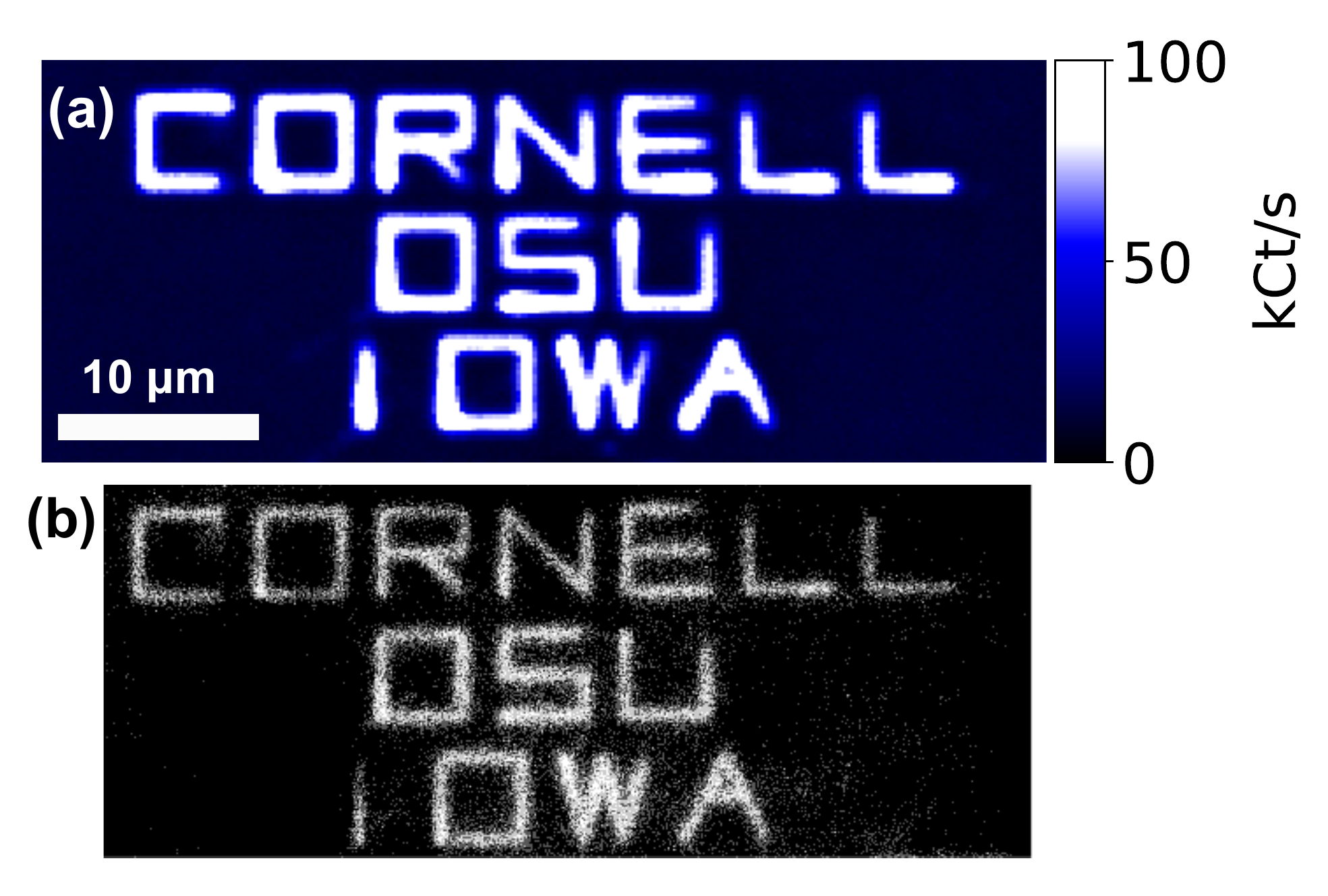}
    \caption{Laser patterning of a 400 nm thick V[TCNE]$_x$ sample (sample 4). (a) Photoluminescence map. High count rate regions are laser damaged, low count rate regions are the remaining undamaged material. (b) Grayscale optical micrograph of the same area. Laser damaged areas are more transparent and appear brighter in the image. This is patterned with 1.7 mW laser power, corresponding to an optical intensity of $\mathrm{1.3\times 10^6~W/cm^2}$.}
    \label{fig:cornell_full}
\end{figure}

Instead of merely being a nuisance, laser damage could also be used for patterning with micron-scale spatial extent. We next explore using laser damage to pattern a V[TCNE]$_x$ film. If thermal degradation is nonlinear with local temperature rise, the attainable feature size could be much smaller than the diffraction limit \cite{Tong2016}.

We show a proof-of-concept demonstration by laser patterning the authors' affiliations on a V[TCNE]$_{x}$ film (Fig. \ref{fig:cornell_full}(a)), where laser written area is damaged, producing a much higher photoluminescence rate. Based on the photoluminescence map and the optical image, the laser patterned feature size is on the order of 1 $\mu$m. Note the photoluminescence map only measures optical properties, and the magnetic properties of laser patterned samples, e.g. magnetization profile, supported spin wave modes, will be the topic of a future study.

\section{Ferromagnetic resonance}

\begin{figure}
    \centering
    \includegraphics[width=0.45\textwidth]{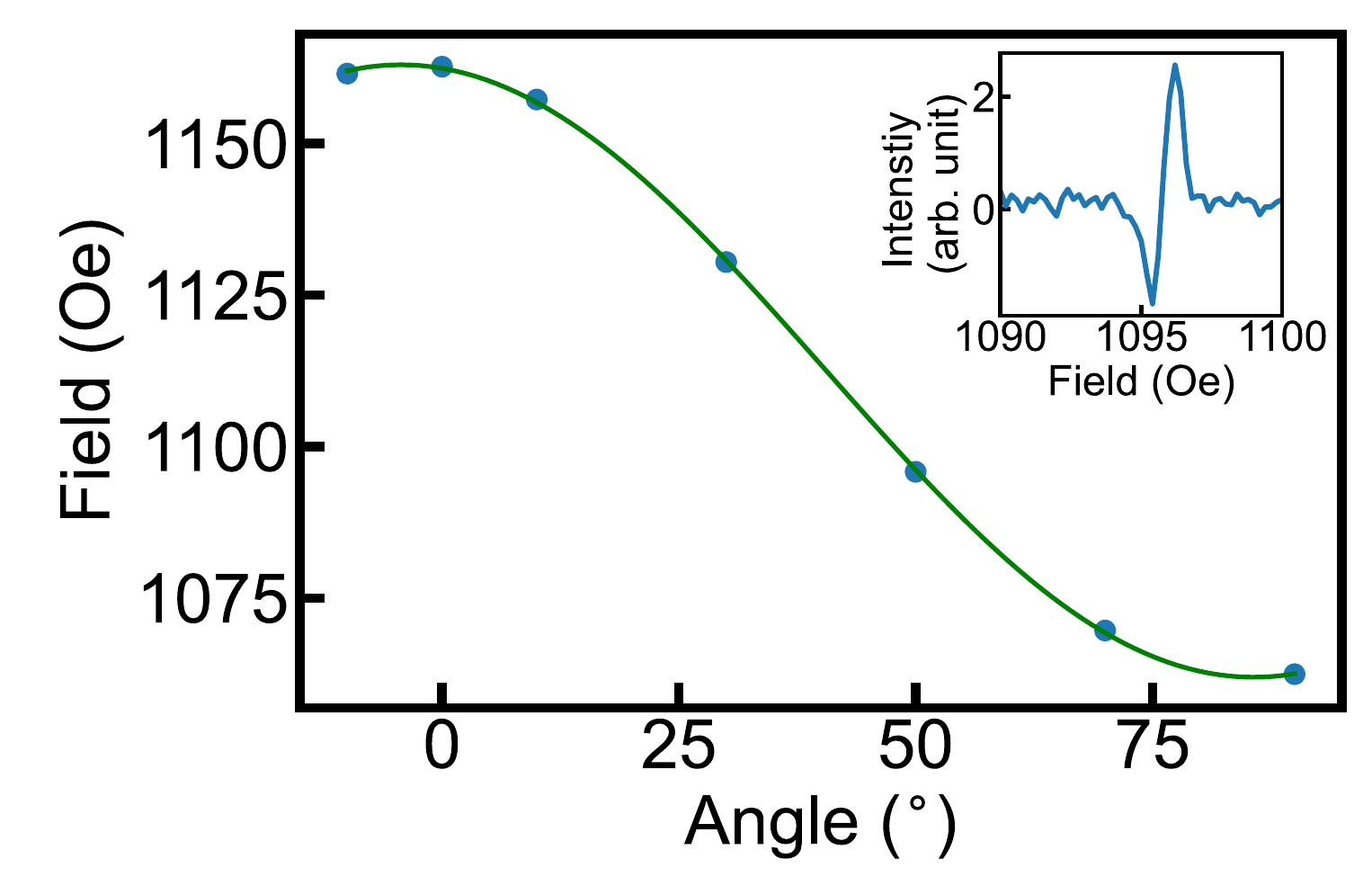}
    \caption{Angle-resolved FMR measured at 3 GHz microwave frequency. Fitted $H_{eff}=\mathrm{68 Oe}$. (Inset) Field sweep at 50$^{\circ}$. Fitted FWHM linewidth is 1.1 Oe. }
    \label{fig:fmr_sweep}
\end{figure}

We next measure the effective magnetization by angle-resolved FMR. The resonance frequency $\omega$ with an external field much greater than the effective field of the material ($H\gg H_{eff}$) is \cite{Yu2014}
\begin{align}
    \omega = \gamma \sqrt{(H-H_{eff}\cos^2\theta)(H-H_{eff} \cos 2\theta)}
\end{align}
where $\gamma$ is the gyromagnetic ratio, $H$ is the external magnetic field, $\theta$ is the angle of the external field with respect to the film normal, $H_{eff}=4\pi M_{eff}=4\pi M_s -H_k$, which is a combination of the saturation magnetization and the anisotropy field. One possible origin of the anisotropy field is from strain due to differential thermal expansion of V[TCNE]$_{x}$ and the underlying substrate \cite{Chilcote2019,Yusuf2020}. In figure \ref{fig:fmr_sweep}, we show a field-modulated FMR signal measured at 3 GHz. Fitting angle-resolved FMR, we extract an effective field $4\pi M_{eff}=\mathrm{68 ~Oe}$. The FWHM linewidth is 1.1 Oe, which is comparable to previously reported values \cite{Franson2019}.

\section{Aging}

\begin{figure}
    \centering
    \includegraphics[width=0.4\textwidth]{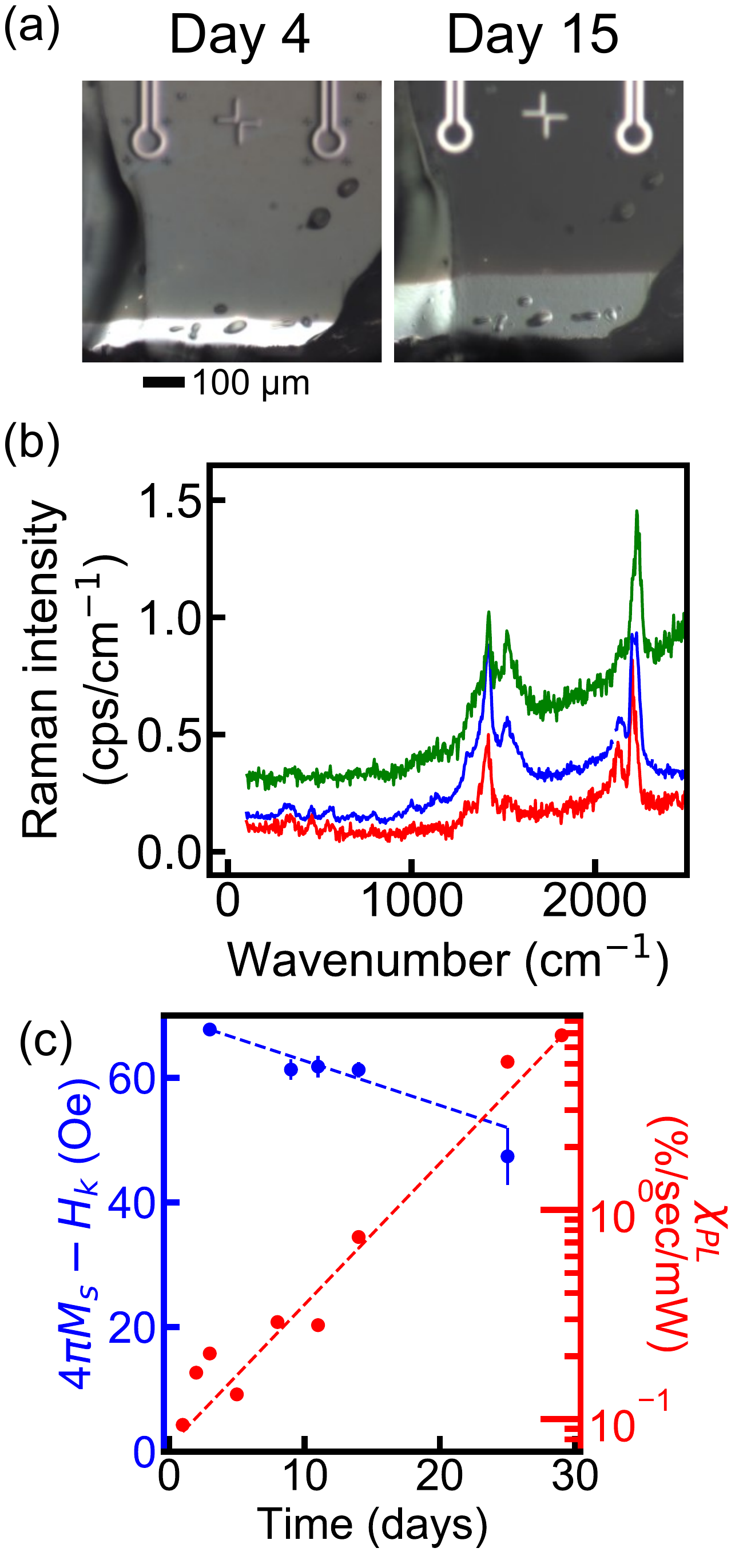}
    \caption{(a) Visual indication of V[TCNE]$_{x}$ aging, showing the aging front advancing from day 4 to day 15. The V[TCNE]$_{x}$ film turns transparent and reveals the underlying reflective aluminum layer. (b) Raw Raman spectrum on day 1 (red), day 32 (blue) and day 89 (green) near sample center. The overall fluorescence/Raman background floor increases and a plateau arises near 1400 $\mathrm{cm^{-1}}$. Over long time (89 days), low wavenumber peaks (300-600 $\mathrm{cm^{-1}}$) and the peak near 2120 $\mathrm{cm^{-1}}$ vanishes. (c) Reduction in $4\pi M_s-H_k$ and increase in laser damage susceptibility. Laser damage susceptibility measured at the center of the sample, using a laser power of 100 - 200 $\mu$W. Fitted lines decay rate is 0.72 Oe/day and laser damage susceptibility increases with a $1/e$ time constant of 6.4 days.}
    \label{fig:sample_B_PL_FMR_degradation_v2}
\end{figure}

Having characterized pristine encapsulated V[TCNE]$_x$, we next monitor its properties as it ages at room temperature in ambient atmosphere. First, we note a visual change in V[TCNE]$_x$, beginning at the sample edge and propagating under the glass coverslip. This material is dark grey-black as deposited. As it ages, a transparency front propagates from the sample edges (Fig. \ref{fig:sample_B_PL_FMR_degradation_v2}(a)). We attribute this chemical change to reaction with diffused oxygen and water across the epoxy encapsulation barrier at sample edges.

In contrast, V[TCNE]$_x$ far from the edges does not show a strong color change with time, However, the lack of an apparent color change does not mean no changes occur. We next characterize sample photoluminescence and Raman spectra at the center of the sample and show how they change in time. Raman spectra show an increase in the fluorescence background and an increase in the Raman intensity near 1300 - 1500 $\mathrm{cm^{-1}}$ as a function of time (Figure \ref{fig:sample_B_PL_FMR_degradation_v2}(b)). In addition, peaks near 1300, 1500 $\mathrm{cm^{-1}}$ increase in Raman intensity compared to the center 1400 $\mathrm{cm^{-1}}$ peak. 
Looking at an even longer timescale (89 days), the low wavenumber peaks (336, 457, 543 $\mathrm{cm^{-1}}$) vanish, and the 2120 $\mathrm{cm^{-1}}$ peak further diminishes. These features are qualitatively similar to those observed in laser damaged samples, with the strongest changes appearing in V--N and $\mathrm{C\equiv N}$ bonding.  This suggests that the slow processes present in room temperature aging are similar to the chemical reactions accelerated by laser damage.

A more drastic signature of aging is an increase in laser damage susceptibility $\chi_{PL}$, which increases exponentially in time as the sample ages (Fig. \ref{fig:sample_B_PL_FMR_degradation_v2}(c)). Concurrently, the effective magnetization $H_{eff}=4\pi M_{eff}$, as measured from angle-resolved FMR, decreases over time (Fig.~\ref{fig:sample_B_PL_FMR_degradation_v2}(c)). This establishes a link between optical properties and magnetic properties, indicating optical measurements could be a local probe of V[TCNE]$_{x}$ film quality.

One limitation of the angle-resolved FMR is that it only measures an intensive quantity $M_{eff}$, which is a sum of shape anisotropy and other anisotropy, e.g. strain induced anisotropy~\cite{Yusuf2020}. Moreover, it is not sensitive to the total magnetic moment of the sample. To disentangle these contributions and understand the total reduction of magnetic material, we monitor aging of another 400 nm thick V[TCNE]$_x$ film (sample 3) with SQUID magnetometry and FMR. 

Here we compare 3 quantities, total magnetic moment $m_{tot}$ measured by SQUID magnetometry, effective magnetization ($4\pi M_{eff}$) and weighted total moment $m_{w}$ measured by angle-resolved FMR. Since $m_{tot}$ and $m_{w}$ are extensive quantities while $4\pi M_{eff}$ is an intensive quantity, we explain below in detail the measurement procedure for each quantity and how we compare them.

\begin{figure}
    \centering
    \includegraphics[width=0.45\textwidth]{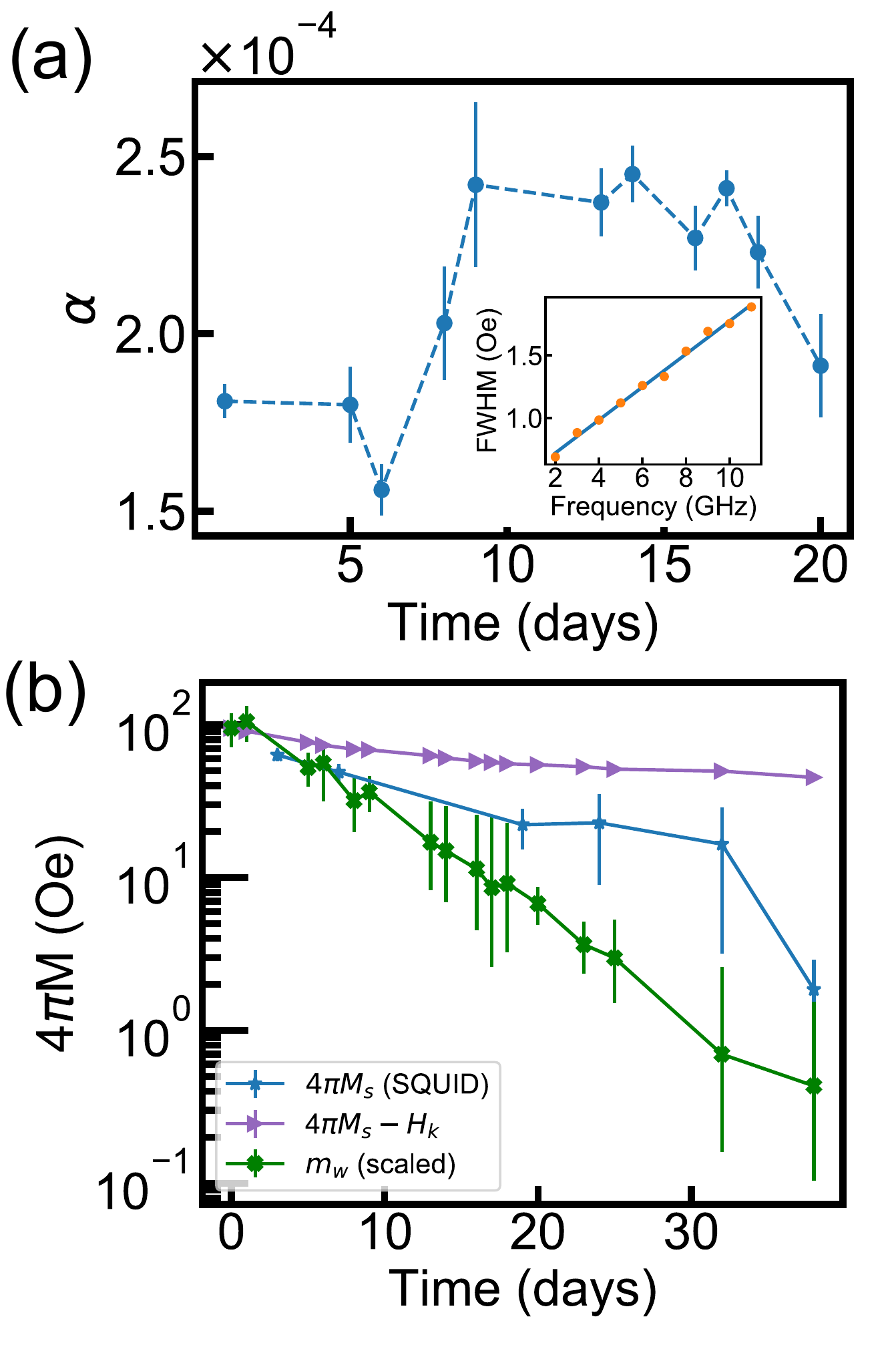}
    \caption{Magnetic properties of an encapsulated V[TCNE]$_{x}$ film as it ages under ambient conditions. (a) The Gilbert damping parameter $\alpha$ extracted from FMR measurements in the range 2-11 GHz with an out of plane DC magnetic field. (Inset) Gilbert damping fit on day 1. $\alpha= (1.8\pm 0.1)\times 10^{-4}$ (b) Magnetic measurements vs. sample aging time. SQUID magnetometry measurements and FMR intensity consistently show a large decrease in magnetic moment. In comparison, $4\pi M_{eff}=4\pi M_s-H_k$ shows a much smaller change over the same period of time.
    }
    \label{fig:dec_sample_B_MPMS_main}
\end{figure}

We compute saturation magnetization from SQUID magnetometry data by normalizing the total moment to the V[TCNE]$_x$ film volume. The volume has an uncertainty of up to 20\%, which is dominated by the thickness uncertainty.  This study lacks a direct measure of the sample thickness for each sample, and there is growth-to-growth variation in the nominal deposition rate as well as variation of growth rate across different positions in the growth chamber. Nonetheless, we are most interested in how $M_s$ changes in the aging process, which is less dependent on initial sample volume.  
We also note that the volume of ferrimagnetic V[TCNE]$_x$ is not constant as the sample ages due to the oxidation front that propagates from sample edge to center (see Fig. \ref{fig:sample_B_PL_FMR_degradation_v2}(a)), resulting in a 33\% reduction in opaque area over 32 days. In figure \ref{fig:dec_sample_B_MPMS_main}(b), we plot the moment normalized to the initial V[TCNE]$_x$ volume. We denote this quantity as $4\pi M_s$ and we interpret it as a scaled total ferrimagnetic moment. 

Next we discuss FMR measurements. We note the sample has a low loss fraction throughout the course of the experiment, with a Gilbert damping varies from $\mathrm{1.5\times 10^{-4}}$ to $\mathrm{2.5 \times 10^{-4}}$ from day 1 to day 20 (Fig.~\ref{fig:dec_sample_B_MPMS_main}(a)). We can unambiguously detect a resonance line shape and the typical uncertainty in $4\pi M_{eff}$ is near 1 Oe. Because we measure FMR with field modulation, we are insensitive to high damping magnetic material, which will appear as a broad background.

This weighted moment $m_{w}$ is computed based on FMR signal strength. With a microwave drive below saturation and a low modulation field, the double integrated signal is proportional to total magnetic moment \cite{Poole1967,Wien2010}. For an absorption derivative line shape $L_{abs}'(H)=a\frac{\Delta H^3(H-H_0)}{(\Delta H^2 + 4 (H-H_0)^2)^2}$, this double integrated signal is proportional to the product of signal amplitude and FWHM $a\Delta H$. We normalize $a \Delta H$ to microwave power and modulation amplitude (see SI) to compute $m_{w}$, which is proportional to the magnetic moment. Because the sample (2450 $\mu$m) is several times wider than the microwave waveguide (430 $\mu$m), FMR sensitivity is non-uniform across the sample. Hence we note $m_{w}$ is a measure of weighted magnetic moment and is more sensitive to the sample portion closer to the microwave waveguide. Moreover, the sample has been remounted multiple times as we alternate between the SQUID and FMR setup, so the coupling between the V[TCNE]$_x$ film and the microwave waveguide varies, which contributes an uncertainty of nearly a factor of 2. With the above subtleties pointed out, we use FMR intensity $m_{w}$ to estimate weighted magnetic moment with low damping.  

Over the course of the study (38 days), $4\pi M_{eff}$ and $m_{w}$ 
reduces by nearly 2 orders of magnitude (Fig.~\ref{fig:dec_sample_B_MPMS_main}(b)). In comparison, $4\pi M_{eff}$ only changes from $95\pm 2$ Oe to $45\pm 1$ Oe. These findings suggest an aging process where the magnetic portion shrinks while remaining low damping. This is consistent with an aging front propagating from sample edge to center, where the center is relatively pristine. These measurements suggest intrinsic aging due to internal chemical reaction at room temperature does not increase damping significantly over 20 days.

To summarize the results of this section over several observations, we  find a qualitative correlation between magnetic properties and optical properties of V[TCNE]$_x$. Namely, a decrease in $4\pi M_{eff}$ correlates with the disappearance of Raman features (300-600 $\mathrm{cm^{-1}}$) involving V--N bonds and an increase in laser damage susceptibility.  DC magnetometry, in combination with FMR, reveals non-uniform aging resulting in a large decrease in total magnetic moment.
The combination of optical and magnetic measurements allow a comprehensive study of V[TCNE]$_x$ properties, relating change in chemical properties and magnetic properties.   

\section{Conclusion}

We characterize pristine encapsulated V[TCNE]$_{x}$ thin films using confocal microscopy, micro-focused Raman spectroscopy, FMR and SQUID magnetometry. Through comparison with \textit{ab initio} calculations, we associate the experimentally observed Raman peaks with particular $\mathrm{C\equiv N,C=C}$ stretching modes. We measure how the sample photoluminescence depends on laser power and observe that laser damage susceptibility has a nonlinear dependence on laser power, which is consistent with a heating based laser damage mechanism. We further explore laser damage as a means of patterning. Studying the spatial profile of laser damaged V[TCNE]$_x$ and their spin waves mode is a subject for future studies.

We identify changes in Raman features and laser damage susceptibility as sample ages. Low wavenumber (300-600 $\mathrm{Cm^{-1}}$) Raman features associated with V--N bonds vanish as sample ages, suggesting changes in bonding between vanadium and TCNE. These findings show that optical measurement is a local probe of V[TCNE]$_{x}$ film magnetic quality and could assess magnetic microstructure quality. 

The existence of a narrow FMR response in encapsulated V[TCNE]$_x$ over 20 days under ambient conditions suggests that intrinsic aging (e.g. not oxidation) does not increase damping over this time interval. This is promising for coherent magnonics using V[TCNE]$_x$ microstructures that are positioned far from encapsulation edges. For quantum applications that use V[TCNE]$_x$ at cryogenic temperatures, one expects that the damping properties will be preserved over an even longer timescale.

\section{Acknowledgements}
Optical Raman experiments, optical writing experiments, waveguide FMR studies of aging, first principles theory, and sample growth of aging study samples were supported by the US Department of Energy Office of Science, Office of Basic Energy Sciences under Award DE-SC0019250.  
Control sample growth and cavity FMR characterization, as well as growth of 2--4 year old samples were supported by NSF DMR-1808704 and DMR-1507775. We acknowledge use of the facilities of the Cornell Center for Materials Research, which is supported through the NSF MRSEC program (DMR-1719875), and the Cornell NanoScale Facility, which is a member of the National Nanotechnology Coordinated Infrastructure and supported the NSF (NNCI-2025233).
We acknowledge helpful discussions with Brendan McCullian.
We thank Hong Tang and Na Zhu for providing 2--4 year old V[TCNE]$_x$ reference samples.



\bibliographystyle{unsrt}
\bibliography{Ref.bib}

\newpage

\section{Supplemental Information}

\subsection{Confocal measurement}

V[TCNE]$_{x}$ samples are measured in a homebuilt confocal microscope with a 532 nm continuous wave laser filtered with a bandpass filter (Iridian 532 BPF, ZX000163). It is focused with a 50x, 0.7 NA objective (Olympus LCPLFLN50xLCD) on the sample. Reflection and fluorescence  from the sample is collected and then filtered by a long pass filter (Iridian 532 LPF nano cutoff, ZX000850). This light is split between a silicon avalanche photodiode (Excelitas SPCM-AQRH-13-FC) and a spectrometer (Princeton Acton SP-2500, focal length 500 nm with a 300 g/mm grating) with a low dark count camera (PyLoN 100 BR).

An incident laser beam of 5 mm $1/e^2$ diameter illuminates the objective back aperture (5 mm diameter). At this truncation ratio, the diffraction limited $1/e^2$ beam diameter is 700 nm~\cite{Urey2004}. At a laser power 500 $\mu$W, peak laser intensity is $\mathrm{2.6\times 10^5~ W/cm^2}$.

\subsection{Raman spectroscopy}

The spectrometer is calibrated with IntelliCal wavelength source and intensity source \cite{IntelliCal2012}. Absolute Raman spectra wavelength accuracy is limited by excitation laser wavelength accuracy, which limits the Raman spectroscopy accuracy to 4-8 $\mathrm{cm^{-1}}$.

For samples that are more susceptible to laser damage, we reduce laser power and raster scan across $\sim$100 $\mathrm{\mu m^2}$ to reduce accumulated laser damage.

We remove cosmic rays from Raman spectra by removing data points that have counts higher than its neighbor above a threshold. The Raman baseline is fitt with asymmetrically reweighted penalized least squares smoothing \cite{Baek2015} with the implementation in the Python module Rampy \cite{RamPy}. Processed Raman spectra are fitted with Lorentzian peaks.

\subsection{\textit{Ab initio} calculations}

The electronic structure and phonon modes of V[TCNE]$_x$ are calculated using the Vienna ab initio Simulation Package (VASP) (version 5.4.4) , which uses a plane-wave basis and pseudopotentials~\cite{Kresse1993,Kresse1994,Kresse1996,KressePRB1996}. The pseudopotentials we used are default options from VASP's official PAW potential set, with 5 valance electrons per V, 4 valance electrons per C, and 5 valance electrons per N atom \cite{KresseJPCondensMatt1994,Kresse1999}.

These pseudopotentials use the generalized gradient approximation (GGA) of Perdew, Burke, and Ernzerhof (PBE) \cite{Perdew1996}. A PBE+U approach with U=4.19 was used in the phonon calculation; with U determined via a linear response method \cite{Cococcioni2005}. This approach is chosen as simple GGA calculations fail to capture the d-orbital behavior of the V atom in our electronic structure calculations. Hybrid functional Heyd–Scuseria–Ernzerhof (HSE) \cite{Heyd2003} with standard range separation parameter $\omega$=0.2 was also tested for this system and showed consistent results with this PBE+U approach. For the rest of the calculation, we used 400 eV for the energy cutoff, and a $\Gamma$-centered $2\times 2\times 2$ k-mesh sampling.

The geometry of V[TCNE]$_x$ is consistent with the work of De Fusco \emph{et al.} \cite{Fusco2009}. The unit cell has a triclinic structure and consists of 1 V atom, 12 C atoms and 8 N atoms. The structure used in the phonon calculation is our own relaxed system under the same set of parameters.

\subsection{Ferromagnetic resonance}

Sample 1 is placed on a 50 $\mathrm{\mu m}$ inner diameter loop antenna. A microwave signal is applied to the antenna with a signal generator (Anristu MG3692C). The reflected signal is routed by a circulator (Narda-MITEQ Model 4923, 2 - 4 GHz) to a diode detector (Herotek DHM185AB). We sweep the magnetic field at a fixed microwave frequency (3 GHz typical) and modulate an external field at a modulation frequency of 587 Hz. Detected power is fed into a lock in amplifier (Signal Recovery 7265) and detected at the modulation frequency.

Sample 3 is placed on a microstrip test board (Southwest Microwave B4003-8M-50). We launch microwave signal at one end and do lock in detection of the transmitted microwave power.

As the sample ages, we increase modulation field amplitude and microwave power to maintain a large enough signal to noise.

\subsection{FMR data analysis}

The spectra are fitted with sums of Lorentzian derivatives, each having an absorptive and a dispersive part \cite{Kalarickal2006}.

\begin{align}
    L_{abs}'(H)&=a\frac{\Delta H^3(H-H_0)}{(\Delta H^2 + 4 (H-H_0)^2)^2}\\
    L_{disp}'(H)&=d\frac{\Delta H^2(\Delta H^2 -4(H-H_0)^2)}{(\Delta H^2 + 4 (H-H_0)^2)^2}
\end{align}
where $\Delta H$ is the full-width at half-maximum of the Lorentzian. 

In angle-resolved FMR, we fit with the following function, using $\gamma,H_{eff},\theta_0$ set as free parameters.

\begin{align}
    \omega = \gamma \sqrt{(H-H_{eff}\cos^2(\theta-\theta_0))(H-H_{eff} \cos 2(\theta-\theta_0))}
\end{align}
where $\gamma$ is the gyromagnetic ratio, $H_{eff}=4\pi M_{eff}$ is the effective magnetization, and $\theta_0$ is a constant offset between the real film normal with respect to the nominal film normal.

Next we explain the procedure of extracting the total magnetic moment from the FMR signal strength. The change in waveguide transmission parameter on magnetic resonance is \cite{Edwards2017,Silva2016}
\begin{align}
    \Delta S_{21}(H_{res}) \propto \frac{\gamma \mu_0 M_s l d_m}{8 Z_0\alpha w}
\end{align}
where $l$ is the film length along the transmission line, $d_m$ is the film thickness, $Z_0$ is the characteristic impedance of the transmission line, and $\alpha$ is the Gilbert damping. At a fixed resonance field, the absorption curve FWHM is $\Delta H_{fwhm}=2\alpha H_{res}$. Hence the integrated area under the absorption curve with respect to external $H$ field is proportional to a weighted magnetic moment, independent of damping.

From the absorption derivative term $L_{abs}'(H)=a\frac{\Delta H^3(H-H_0)}{(\Delta H^2 + 4 (H-H_0)^2)^2}$, we extract an equivalent area $a H_0$ as an estimate of the total magnetic moment.

We adjust for microwave power by normalizing the detected signal to microwave power ($P_0$) and the power-to-voltage conversion factor ($K$(V/mW)) of the diode detector. Namely, the normalized signal is
\begin{align}
    \frac{V_{sig}}{P_0 K}
\end{align}

In modulation detection, when modulation field is much smaller than linewidth, the lock in detected signal is proportional to a derivative of a Lorentzian absorption. When modulation field is comparable to linewidth, detected signal is no longer a simple derivative and we correct for that by modeling the line shape.

Note the FMR line shape has both an absorptive derivative and a dispersive derivative. Here we only use the absorptive derivative to extract the total magnetic moment.

\subsection{Laser heating}\label{app:laser heating}

In this section, we estimate local temperature rise due to laser heating of a V[TCNE]$_x$ film grown on a glass substrate. We approximate the glass substrate thermal conductivity to be $\kappa_1=1.4 ~W m^{-1} K^{-1}$. We approximate V[TCNE]$_x$ and epoxy to have a similar thermal conductivity $\kappa_2=0.3~ W m^{-1} K^{-1}$ \cite{Lee1982,Li2018}.

Because the laser spot size (350 nm) is much smaller than the epoxy thickness (5-10 $\mu$m), we model both the substrate and V[TCNE]$_{x}$ plus epoxy as semi-infinite. The temperature rise profile for a point heat source at the interface is
\begin{align}
    T(r) = \frac{1}{4\pi \bar{\kappa}} \frac{\dot{Q}}{r}
\end{align}
where $\bar{\kappa} = (\kappa_1+\kappa_2)/2$, assuming temperature is fixed at 0 at infinity.

We model the actual heat source as a surface Gaussian heat source 
\begin{align}
    \frac{2P_0}{\pi w^2 }\exp\bigg(-\frac{2r^2}{w^2}\bigg)
\end{align}
where $w$ is the $1/e^2$ beam radius.

The maximum temperature rise occurs at the interface along the beam center,
\begin{align}
    T_{max}=\frac{1}{2\sqrt{2\pi} \bar{\kappa}}\frac{ P_0}{w}
\end{align}

\noindent At 0.7 NA, $1/e^2$ beam radius $w\approx 350~ \mathrm{nm}$. 100 $\mu$W of power causes a local temperature rise of 70 K.

A more realistic estimate takes into account of the finite absorption depth of V[TCNE]$_{x}$, which is of the order 100-400 nm. As this is comparable to the beam size, the peak temperature rise is of the same order of magnitude.

\subsection{Long term aging}

Visual inspection provides preliminary information on V[TCNE]$_x$ film quality, but it is neither a quantitative nor a definitive measure. While discolored V[TCNE]$_{x}$ is almost certainly aged, even fully opaque film can be magnetically inactive\cite{Froning2015}.

To better understand long term degradation, we examine 2 other V[TCNE]$_{x}$ samples. One is a 4 year old uniform film, the other is a 2 year old, patterned 100 $\mu m$ wide bars. Both of them are nominal 1 $\mu$m thick, estimated from growth time. Both of them are visually opaque and yet are magnetically inactive. 

\begin{figure*}
    \centering
    \includegraphics[width=0.9\textwidth]{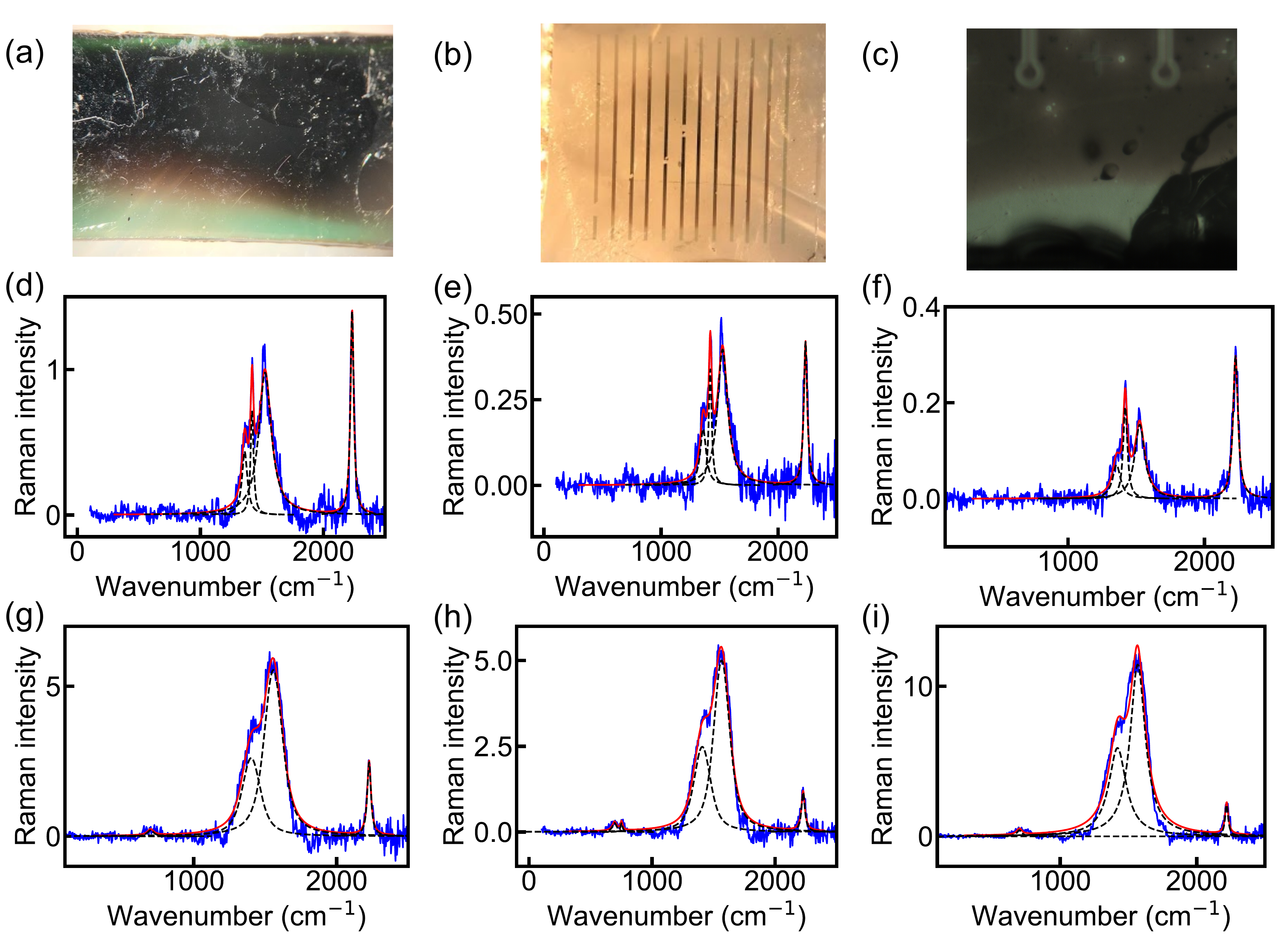}
    \caption{Long term study of encapsulated V[TCNE]$_{x}$ Raman spectra. The Raman spectra are baseline subtracted. We probe the center opaque areas away from encapsulation edges in all 3 samples. (a) 4 year old, 1 $\mu$m thick uniform film. (b) 2 year old, 1 $\mu$m thick, patterned 100 $\mu$m wide bar sample. (c) (Sample 1 in the main text) 89 days old, 400 nm thick film with aluminum encapsulation.  Note that the V[TCNE]$_{x}$ film has been remounted on the antenna and their relative position differ from that in the main text. (d)-(f) The laser is raster scanned across near a 50$\mu m^2$ area to reduce laser damage. Laser power is (d) 190 $\mu$W, (e) 190 $\mu$W, (f) 51 $\mu$W .  (g)-(i) The laser power is 1.9 mW with the laser focused at one spot to evaluate the effect of laser damage.}
    \label{fig:long_term_study}
\end{figure*}

The samples show qualitatively similar peaks (Fig. \ref{fig:long_term_study}), with 3 peaks near 1300-1500 $\mathrm{cm^{-1}}$ and a single peak near 2200 $\mathrm{cm^{-1}}$. Under strong laser illumination, 3 peaks 1300-1500 $\mathrm{cm^{-1}}$ merge into 2 peaks, and become brighter than the 2200 $\mathrm{cm^{-1}}$ peak. Crucially, the spectra differ significantly from aged V[TCNE]$_{x}$. Degraded samples don't have low wavenumber peaks (300-600 $\mathrm{cm^{-1}}$) and the small 2120 $\mathrm{cm^{-1}}$ peak as seen in pristine samples, further supporting a link between these fine features with good magnetic properties.

\subsection{SQUID magnetometry}

We measure a 400 nm thick V[TCNE]$_x$ film (sample 3) in a Quantum Design MPMS 3 SQUID magnetometer in the vibrating sample magnetometer (VSM) mode. The sample is mounted to a quartz holder with a small dab of GE varnish. We measure moment vs. field at 300 K in a four-quadrant sweep, from --20,000 Oe to +20,000 Oe back to --20,000 Oe. We use a palladium reference to correct for the absolute field error~\cite{Pd2018}. The applied field is in the plane of the V[TCNE]$_x$ film.

We normalize the measured magnetic moment to the initial film volume, which is estimated from the opaque film area (7.27 $\mathrm{mm^2}$) and the nominal film thickness 400 nm.

\begin{figure*}
    \centering
    \includegraphics[width=0.8\textwidth]{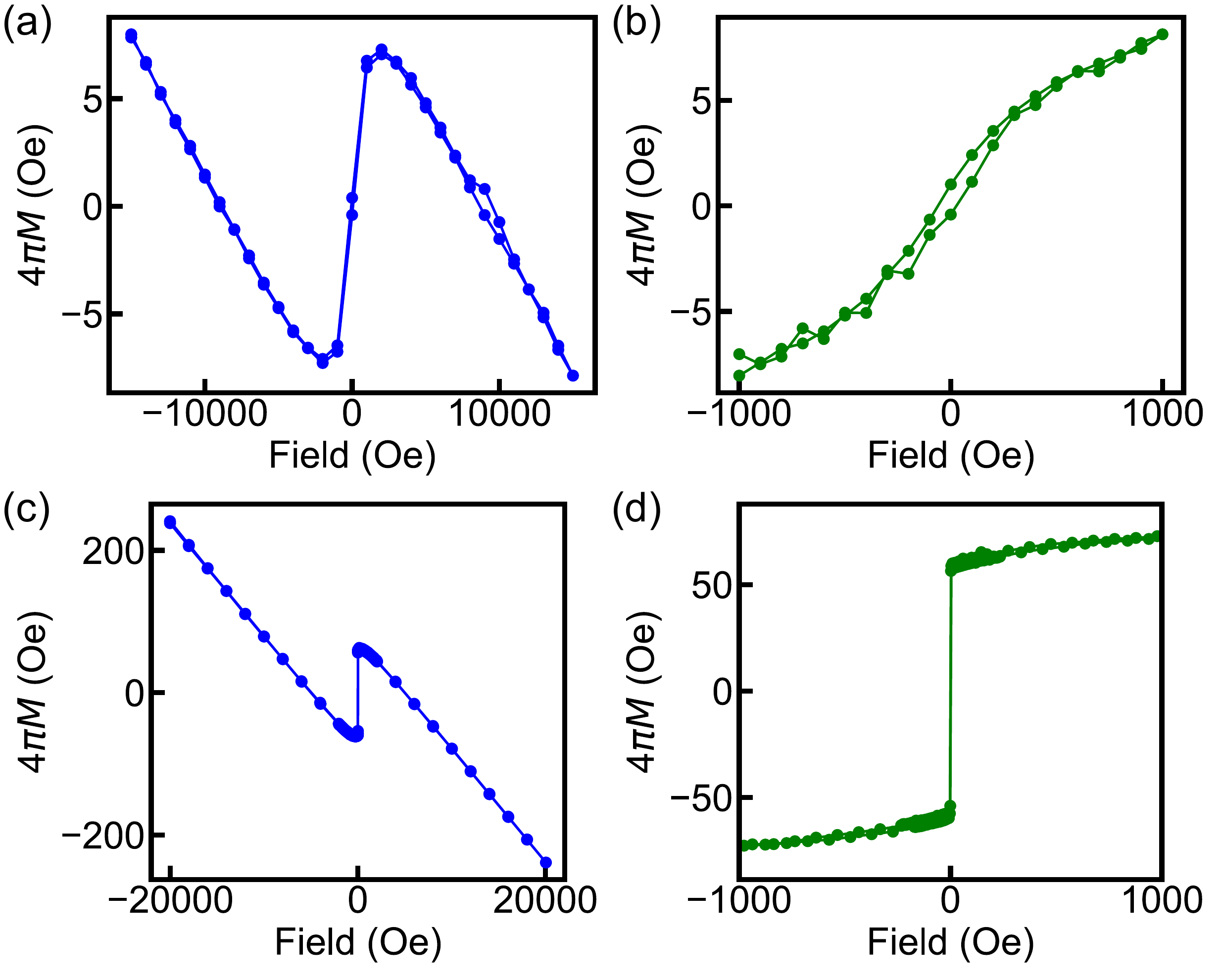}
    \caption{Magnetization $4\pi M$ vs. field for GE varnish and sample 3. (a)-(b) Quartz sample holder and GE varnish response. Measured moment is normalized to initial V[TCNE]$_x$ volume. (a) Total magnetization (b) Linear background subtracted magnetization. (c)-(d) V[TCNE]$_x$ sample on day 3. (c) Total magnetization (d) Linear background subtracted magnetization.}
    \label{fig:dec_sample_B_all}
\end{figure*}

The magnetization vs. field data has a large diamagnetic background (Fig.~\ref{fig:dec_sample_B_all}(c)), which we subtracted away by fitting to a linear background at fields $|H|>$ 2000 Oe. We also measure the magnetic response GE varnish, which has a small diamagnetic and ferromagnetic component (Fig.~\ref{fig:dec_sample_B_all}(a,b)). The magnetic moment of GE varnish is normalized to the same initial V[TCNE]$_x$ volume. Note that this is an equivalent V[TCNE]$_x$ magnetization error due to GE varnish background but not the physical GE varnish magnetization. As the amount of GE varnish applied in different runs varies, one cannot simply subtract a fixed background from the data. We note GE varnish has a gradual moment vs. field dependence over --1000 Oe to +1000 Oe (Fig.~\ref{fig:dec_sample_B_all}(b)). Because the applied field is in plane, V[TCNE]$_x$ should be close to fully magnetized at 200 Oe. Therefore, we estimate V[TCNE]$_x$ $4\pi M_s$ to be the magnetization at 200 Oe, where background magnetization is small. We estimate the uncertainty as the difference between magnetization at 200 Oe and 500 Oe.

\end{document}